\documentclass[useAMS,usenatbib]{mn2e} 
\usepackage{times}

\def\spose#1{\hbox to 0pt{#1\hss}} 
\newcommand\lsim{\mathrel{\spose{\lower 3pt\hbox{$\mathchar"218$}} 
     \raise 2.0pt\hbox{$\mathchar"13C$}}} 
\newcommand\gsim{\mathrel{\spose{\lower 3pt\hbox{$\mathchar"218$}} 
     \raise 2.0pt\hbox{$\mathchar"13E$}}} 
 
\title[]{Probing  the existence  of the  $E_{\rm  peak}$-$E_{\rm iso}$ 
correlation in long Gamma Ray Bursts} 
 
\author[Ghirlanda G.  et al.]
{Giancarlo Ghirlanda$^{1}$\thanks{E-mail: ghirlanda@merate.mi.astro.it},   
Gabriele   Ghisellini$^{1}$,  Claudio Firmani$^{1,2}$ \\  
$^{1}$Osservatorio Astronomico  di Brera, via  E.Bianchi 46,  I-23807 Merate, Italy\\  
$^{2}$Instituto de Astronom\'{\i}a,  U.N.A.M.,  A.P.    70-264,  
 04510,  M\'exico,  D.F., M\'exico} 
\begin{document} 
 
\date{} 
 
\pagerange{\pageref{firstpage}--\pageref{lastpage}} \pubyear{2002} 
 
\maketitle 
 
\label{firstpage} 
 
\begin{abstract} 
We probe the existence of the $E_{\rm peak}$--$E_{\rm iso}$
correlation in long GRBs using a sample of 442 BATSE bursts with known
$E_{\rm peak}$ and with redshift estimated through the lag--luminosity
correlation. This sample confirms that the rest frame peak energy is
correlated with the isotropic equivalent energy.  The distribution of
the scatter of the points around the best fitting line is similar to
that obtained with the 27 bursts with spectroscopic redshifts.
We interpret the scatter in the $E_{\rm peak}-E_{\rm iso}$ plane as
due to the opening angle distribution of GRB jets.  By assuming that
the collimation corrected energy correlates with $E_{\rm peak}$ we can
derive the observed distribution of the  jet opening angles,
which turns out to be log--normal with a peak value of $\sim
6.5^\circ$.
\end{abstract} 
 
 
\section{Introduction} 
 
Since the discovery of the cosmological distance scale to Gamma Ray
Bursts in 1997 (Djorgovski et al. 1997) several GRB redshifts have
been spectroscopically
measured\footnote{http://www.mpe.mpg.de/$\sim$jcg/grbgen.html for a
continuously updated table.}.  The small sample of bursts, now
comprising 42 events, allowed in the latest years the investigation of
the GRB rest frame properties.  Several correlations have been
discovered between GRBs spectral properties:
\begin{enumerate} 
\item  
the  spectral lag - luminosity  correlation $\tau$--$L_{\rm iso}$ 
(Norris, Marani \& Bonnel  2000; Norris 2002) 
\item   
the  variability  -  luminosity correlation  $V$--$L_{\rm  iso}$ 
(Fenimore \& Ramirez--Ruiz 2000; Reichart et al.  2001). 
\item  
the peak energy - isotropic energy correlation $E_{\rm peak}$--$E_{\rm 
iso}$ (Amati et al. 2002; Lloyd \& Ramirez-Ruiz 2002). We will 
call it the Amati correlation. 
\item  
the  peak energy  - isotropic peak luminosity correlation  $E_{\rm peak}$--$L_{\rm 
iso,peak}$ (Yonetoku et al. 2004). 
\item   
the peak energy - collimation corrected energy correlation  
$E_{\rm peak}$-$E_{\gamma}$ (Ghirlanda, Ghisellini, Lazzati 2004, GGL04 
hereafter).  We will call it the Ghirlanda correlation. 
\end{enumerate} 
 
All these  correlations can be represented by  powerlaws: their slopes
are summarized in Tab. 1.  Note that, for the last three correlations,
the  peak energy  $E_{\rm  peak}$  of the  $\nu  F_{\nu}$ spectrum  is
obtained by  fitting the GRB  spectrum {\it time--integrated  over the
burst  duration.}   
 
Despite  several  attempts to  explain  these  correlations (see  e.g.
Eichler D.  \& Lenvison A.  2004;  Liang E., Dai Z.  \& Wu X.F. 2004),
there is no universally accepted interpretation yet.  One of the major
drawback of  these correlations  is that they  were found for  a small
sample  of  bursts, mainly  due  to  the  limited number  of  reliable
redshift measurements.  However, one  test which can be performed even
without knowing  the redshift is to  check if GRBs  with known fluence
and  $E_{\rm  peak}$   are  consistent  with  --  say   --  the  Amati
correlation, by considering all possible redshifts.

This kind of consistency check has been performed very recently by two
groups which tested the Amati and the Ghirlanda correlation.  Nakar \&
Piran  (2004, hereafter  NP04)  tested the  Amati  correlation in  the
observed  plane  $E_{\rm  peak,obs}$--$F$  (where  $F$  is  the  burst
observed fluence) using a sample  of bright BATSE bursts (from Band et
al.   1993;  Jimenez,  Band  \&  Piran  2000).   They  concluded  that
$\sim$40\%  of the BATSE  long (bright)  bursts are  inconsistent with
this relation.
 
Band \& Preece (2005, hereafter BP05) extended this work with a larger
sample  of BATSE  GRBs (760  bursts from  Mallozzi et  al.   1998) and
tested both the Amati and  the Ghirlanda correlation.  They claim that
at  least  88\%  of  their  bursts are  inconsistent  with  the  Amati
correlation and 1.6\% with the Ghirlanda correlation.  They also claim
that these conclusions are robust, since they are model--independent.
 
An important point to notice is what is meant by Amati correlation.
Originally, Amati et al.  (2002) found $E_{\rm peak}\propto E_{\rm
iso}^{0.5}$ with a small scatter, but using only 9 bursts.  Later,
GGL04 (see their Tab. 1 and Tab. 2) collected 23 GRBs (including the
above 9) with redshift measurements and published peak energy, finding
a slightly different slope and, more importantly, a larger scatter
around the correlation, whose significance was nevertheless improved
(due to the increased number of points).  In the same paper, it was
demonstrated that for all bursts with a good measurement of the jet
opening angle (through the detection of a break in the light curve of
the afterglow), the scatter of the Amati correlation was entirely due
to the different opening angles of the jet: bursts with the same
collimation--corrected energy $E_{\gamma}$ and the same intrinsic
$E_{\rm peak}$ have different $E_{\rm iso}$ because they have
different jet opening angles.  The scatter of the Amati correlation,
far from being due, i.e.  to ill--measured quantities, has a physical
origin and is revealing of the distribution of the opening angles of
the jets.
 
Bearing this in mind, in this letter we test if $E_{\rm peak}$
correlates with $E_{\rm iso}$ even if we do expect a large scatter
around the best fitting line (if any).  To this aim we use a large
sample of bursts with ``pseudo redshifts" which have been estimated
through the lag--luminosity relation (Sec. 2).  Having found that
indeed the $E_{\rm peak}$-$E_{\rm iso}$ correlation exists (Sec. 3), we follow GGL04 and interpret the scatter as due to the
distribution of jet opening angles, finding it (Sec. 4).  Finally
(Sec. 5), we discuss why NP04 and BP05, who also use a large sample of
BATSE bursts, arrive to different conclusions.
 
During  the completion  of this  work, we  received a  manuscript from
Bosnjak et al.   By studying the same problem  with an independent and
complementary approach, they arrive  at the same conclusions presented
in this letter.
 
In  the following  we  adopt a  standard  $\Lambda$CDM cosmology  with
$\Omega_{\Lambda}=0.3$, $\Omega_{m}=0.7$ and $h_0=0.7$.

\begin{table} 
\centering 
\begin{tabular}{@{}lllll@{}} 
\hline   
& $L_{\rm  iso,peak}$ &  $E_{\rm iso}$  & $E_{\gamma}$  & Ref \\ 
\hline  
$\tau$         &0.87$^{+0.19}_{-0.13}$                & ....           & ....           & (1), (2) \\  
$V$            &$0.3^{+0.11}_{-0.07}$& ....           & ....           & (3) \\  
$E_{\rm peak}$ &$0.5\pm0.05$         &$0.52\pm 0.06$  &$0.706\pm 0.04$ & (4), (5), (6)\\  
\hline 
\end{tabular} 
\caption{ 
Slopes of the correlations between rest frame  GRB quantities 
(for instance: $\tau \propto L_{\rm  iso,peak}^{0.87}$). 
Ref:   
(1) Norris et  al. 2000; 
(2) This work (Sec.3)
(3) Fenimore \& Ramirez--Ruiz 2000,  Reichart et al.  2001; 
(4) Yonetoku et al. 2004; 
(5) Amati  et al.  2002; 
(6) Ghirlanda,  Ghisellini \& Lazzati 2004. 
} 
\end{table}

\section{The GRB sample} 
 
As described above  the Amati correlation involves the  peak energy of
the time integrated spectrum of the prompt emission.  This is obtained
by fitting the  GRB spectrum time integrated over  the entire duration
of the burst (see e.g. Band et al. 1993).
 
Yonetoku  et  al. (2004;  Y04  hereafter)  have  recently published  a
catalog of 689 long BATSE  GRBs obtained through the spectral analysis
of  the  high spectral  resolution  BATSE  data.   The main  selection
criterium adopted  by Y04  is on the  peak flux (i.e.   $\gsim 2\times
10^{-7}$  erg cm$^{-2}$  s$^{-1}$).   Band, Norris  \& Bonnell  (2004,
BNB04 hereafter) derived for the  BATSE GRBs in the sample of Mallozzi
et al. 1998 (the same used in BP05) their pseudo redshifts through the
lag--luminosity  correlation  (Norris  et  al.   2000).   This  sample
contains 1165 bursts.  We have  selected 578 sources contained both in
Y04 and in BNB04 to have a sample of bursts with a good measurement of
$E_{\rm peak}$ and  an estimate of their redshift.   For each selected
burst, we took the fluence from the current BATSE GRB catalog.
\footnote{http://cossc.gsfc.nasa.gov/batse/BATSE\_Ctlg/index.html}   We 
excluded  from the final  sample the  sources with  particularly large 
uncertainties on  the parameters which are  used for the  test that we 
present in the next sections: 
\begin{itemize} 
\item  21 bursts with $E_{\rm peak} \le 40$ keV, since this value
is  very close  to the  BATSE threshold,  and this  makes  the fitting
procedure highly uncertain (see e.g. Lloyd \& Petrosian 2002);
\item 5  bursts  with  errors in  $E_{\rm  peak}$ greater  than 
$E_{\rm peak}$ itself; 
\item  6 bursts  with errors on  their fluence greater  than their 
fluence.\\
\noindent This  sample contains 546 GRBs  and it will be  used for the
test discussed in Sec.5.\\
In addition, to the aim of testing the Amati correlation with a sample
of GRBs with reliable  redshift estimates we excluded:
\item  
94 bursts with  errors in their estimated lag greater than 
the lag itself. 
\item  
10 bursts with redshift larger than 10. 
\end{itemize} 
The  final  sample contains  442  GRBs and  this  is  used in  Sect.3.
Although the estimated pseudo redshifts should be taken with care when
considering  specific objects,  the  distribution of  redshifts for  a
large sample  of bursts should be  more reliable.  In Fig.   1 we show
that the  distribution of  the 442 pseudo  redshifts of our  sample is
similar  to the distribution  of spectroscopically  measured redshifts
(27  objects).    Even  if  not   complete,  this  sample   should  be
representative of moderately bright BATSE bursts.
\begin{figure} 
\begin{center} 
\vspace{5.5cm} 
\includegraphics{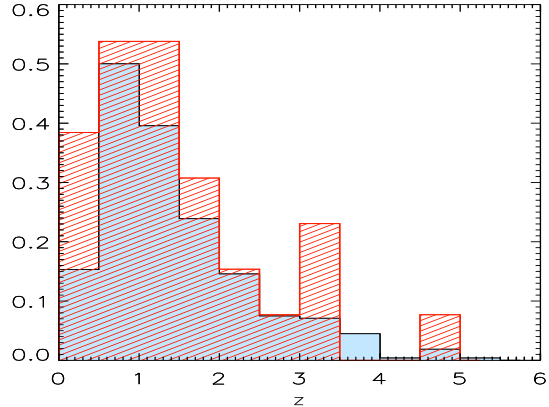} 
\caption{ Redshift  distribution of the  sample of bursts  with pseudo
redshift  (filled histogram,  442 objects),  compared to  the redshift
distribution   of   GRBs    with   measured   spectroscopic   redshift
(red--hatched   histogram,   27   objects).  Distributions   are
normalized.}
\end{center} 
\label{nz} 
\end{figure} 
 
\section{Testing the existence of an $E_{\rm peak}-E_{\rm iso}$ correlation} 
 
Amati et al.  (2002) showed that the rest frame $E_{\rm peak}-E_{\rm
iso}$ are correlated in 9 long GRBs detected by $Beppo$SAX.  They
found a correlation coefficient of 0.949 with a chance probability
$P=5\times 10^{-3}$.  This correlation was confirmed by GGL04 using a
larger sample (objects listed in their Tab. 1 and 2) of 23 GRBs, with
spectroscopically measured redshifts and published peak energy.  Here,
we update this sample with 4 more GRBs whose redshifts have been
recently measured (see Ghirlanda et al. 2005)\footnote{See also
http://www.merate.mi.astro.it/$\sim$ghirla/deep/blink.htm}.  With
these 27 bursts the Spearman rank correlation coefficient is
$r_{s}=0.87$ and a chance probability $P=3.4\times 10^{-9}$.
 
In Fig. 2 we show the rest frame $E_{\rm peak}$--$E_{\rm iso}$ of
these 27 GRBs (red symbols).  The fit with a powerlaw model to
these data points (weighting for the errors on both coordinates) is:
\begin{equation} 
{E_{\rm  peak} \over  {100\, {\rm  keV}}} \, = \, (3.2\pm 0.1) 
{\left(E_{\rm  iso} \over {1.1\times 10^{53} \, {\rm erg}} \right)^{0.56 \pm 0.02}} 
\end{equation} 
with a $\chi^{2}=129.6/25$ dof.  The distribution of the scatter
measured along the fitting line (i.e. the distances of the data points
to the fitting line) of the 27 GRBs is shown in Fig.  3 by the
red-hatched histogram.  Fitting with a Gaussian we find a dispersion
$\sigma=0.22$.

\begin{figure} 
\begin{center} 
\vspace{8cm} \includegraphics{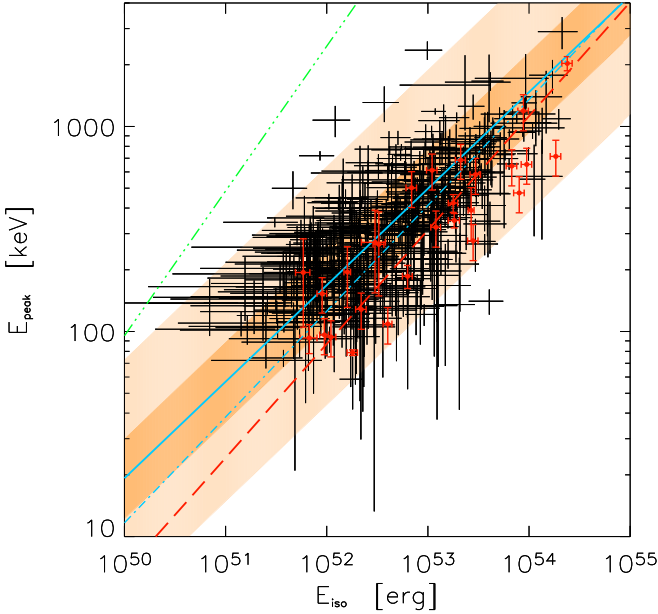} 
\caption{ Rest frame peak energy $E_{\rm peak}$ versus isotropic
energy $E_{\rm iso}$.  Red filled circles are the 27 GRBs from
Tab.  1 and 2 of GGL04 updated with the 4 new events with
spectroscopic redshifts and the long--dashed red line is their best
powerlaw fit.  The black crosses are the 442 GRBs with redshift given
by the lag--luminosity correlation (BNB04). The solid blue line
represents their best powerlaw fit weighted for the errors on both
coordinates. The dot--dashed line is the fit to the complete sample
(442 + 27 GRBs).  The results of these fits are given in the text.
Also shown (triple dot--dashed green line) is the Ghirlanda
correlation.  The shaded regions represent the 1$\sigma$ and 3$\sigma$
width of the dispersion of the black crosses around their correlation,
as represented by the blue solid line.  }
\end{center} 
\label{corr} 
\end{figure} 
 
\begin{figure} 
\begin{center} 
\vspace{7.8cm} 
\includegraphics{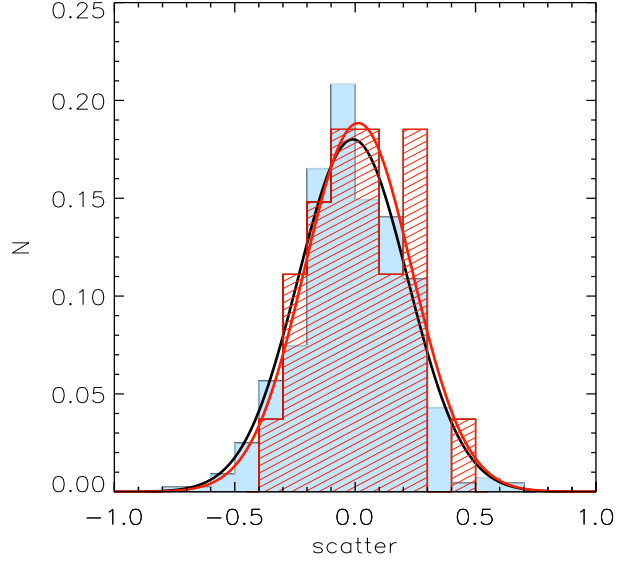} 
\caption{ Solid histogram: Scatter of the 442 GRBs around their
powerlaw fit (Eq. 6). The black solid line is the gaussian fit to this
distribution.  Red hatched histogram: scatter of the 27 GRBs
with spectroscopic redshift around their powerlaw fit (Eq.  1). The
red solid line is the gaussian fit. }
\end{center} 
\label{scatter} 
\end{figure} 
 
For the 442  GRBs with pseudo redshifts, we  have estimated the errors
on the  rest frame $E_{\rm iso}$  and $E_{\rm peak}$  as follows. From
$E_{\rm peak}=E_{\rm peak,obs}(1+z)$ we derive
\begin{equation} 
\left(\frac{\sigma_{E_{\rm peak}}}{E_{\rm peak}}\right)^2= 
\left(\frac{\sigma_{E_{\rm peak,obs}}}{E_{\rm peak,obs}}\right)^2+ 
\left(\frac{\sigma_{z}}{1+z}\right)^2 
\end{equation} 
Similarly from $E_{\rm iso}=4\pi d^2_L  F / (1+z)$, where $d_L$ is the
luminosity distance and $F$ is the GRB bolometric fluence, we obtain:
\begin{equation} 
\left({ \sigma_{E_{\rm iso}} \over E_{\rm iso} }\right)^2= 
\left({ 2 d^\prime_L\, \sigma_{z} \over d_L }- 
{ \sigma_z \over 1+z } \right)^2+ 
\left({ \sigma_F \over F }\right)^2 
\end{equation} 
where $d^\prime_L=\partial d_L/\partial z $.  
 
BNB04 do  not give the error  on the estimated  redshifts. However, it
can be obtained from $L_{\rm iso}=4\pi d^2_L \Phi$, where $\Phi$ is the peak
flux, as follows:
\begin{equation} 
\sigma_{z}^2= \frac{d^2_{L}}{4 {d^\prime_{L}}^2} 
\left[\left(\frac{\sigma_{L_{\rm iso}}}{L_{\rm iso}}\right)^2+ 
\left(\frac{\sigma_{\Phi}}{\Phi}\right)^2\right] 
\end{equation} 

The term  $\sigma_{L_{\rm iso}}/L_{\rm iso}$ can be  obtained from the
lag--luminosity relation  $L_{\rm iso}=10^{B}\tau^n$, using  the error
on  the lag,  i.e.  $\sigma_{\tau}$  given  by BNB04  in their  table.
However, also the error on the  slope $n$ and on the normalization $B$
contribute to the  error on the luminosity.  Norris  et al. (2000) and
BNB04  do not  give the  uncertainty  on these  parameters.  For  this
reason we  recomputed the lag--luminosity  relation (with the  data of
Table  1 of  Norris 2000)  in the  ``barycenter" of  the  data points,
namely  $\tau_{b}=0.036$  s  and  $L_{iso,b}=3.36\times  10^{52}$  erg
s$^{-1}$, in order to treat $B$ and $n$ as uncorrelated.
  
We found  exactly the  same slope and  normalization but we  could now
also    find   the   uncertainties    on   these    quantities,   i.e.
$n\pm\sigma_{n}=1.149\pm0.205$  and (in the  coordinate system  of the
barycenter the  normalization is null  but its error is  nonzero) i.e.
$B\pm\sigma_{B}=0.0\pm 0.116$.  We can finally compute the error:
\begin{equation} 
\left( {\sigma_{L_{\rm iso}}\over L_{\rm iso}} \right)^2= 
2\ln10\left[\left(\sigma_{n}\log\left(\frac{\tau}{\tau_{b}}\right)\right)^{2}+\sigma_{B}^{2}\right] +
\left(n\frac{\sigma_{\tau}}{\tau}\right)
\end{equation} 

Fig.  2 shows the 442 GRBs with pseudo redshifts (black 
symbols)
\footnote{Note that $E_{\rm iso}$ is computed accounting for the band
k-correction as described in Eq.2 of GGL04.}.  Interestingly, we find
that these 442 GRBs still show a highly significant correlation
between the rest frame $E_{\rm peak}$ and $E_{\rm iso}$ with a
Spearman correlation coefficient $r_s= 0.7$ and a chance probability
$P=2.1\times 10^{-65}$. The partial correlation coefficient,
i.e. removing the effect of $z$, is $r_p=0.65$ and, therefore,
indicates that the correlation is not determined by the common
dependence of Eq. 2 and Eq. 3 on the term $(1+z)$.  The fit (weighting
errors on both coordinates) with a powerlaw to these data gives:
\begin{equation} 
{E_{\rm  peak} \over  100\, \,{\rm  keV}} \, =\,  (4.34\pm 0.05)   
\left( {E_{\rm  iso} \over 7.5 \times 10^{52} \, {\rm erg}} \right)^{0.47 \pm 0.01} 
\end{equation}
with a $\chi^{2}=1735/440$ dof.  In this case the scatter distribution
(filled histogram in Fig. 3) has a dispersion $\sigma=0.22$,
consistent with the dispersion of the 27 GRBs with spectroscopic
redshifts.
It is remarkable that both the slopes of the two correlations (Eq. 1
and Eq.  2 above) and their scatter distributions are very similar.
However, the two correlations have different normalizations: the 27
GRB with spectroscopic redshifts seem to define an envelope to the
distribution of all GRBs of this sample.  A possible reason for this
behavior is that GRBs with spectroscopically measured redshifts tend
to have fluences larger than the average, at least when comparing GRBs
detected by BATSE or $Beppo$SAX.  This fact has been pointed out also
by NP04.

We finally consider the total sample of bursts, i.e.  the 27 GRBs with
spectroscopic redshifts plus the 442 with pseudo redshifts, and find a
Spearman correlation coefficient $r_{s}=0.7$ and $P=1.5\times
10^{-71}$, and the partial correlation coefficient is $r_p=0.66$.  The
fit with a powerlaw gives:
\begin{equation} 
{E_{\rm  peak} \over  {100\, \, {\rm  keV}}} \, =\, (3.64\pm0.04)   
\left( {E_{\rm  iso} \over 7.9\times 10^{52} \, {\rm erg} } \right)^{0.51 \pm 0.01} 
\end{equation} 
with a $\chi^{2}=2185/463$ dof.

\section{The origin of the scatter in the $E_{\rm peak}$--$E_{\rm iso}$ plane} 
 
Both the 27 GRBs with spectroscopic redshifts and the 442 GRBs
here considered have a relatively large scatter in the $E_{\rm
peak}$--$E_{\rm iso}$ plane.  In GGL04 it has been shown that for the
few GRBs with measured $\theta_{\rm jet}$ this scatter reduces to only
0.1 dex on the Ghirlanda correlation.  This suggests (see also Firmani
et al. 2005) that the scatter might be due to the angle distribution
of GRBs.

If we make the hypothesis that the Ghirlanda correlation has a scatter 
much smaller than the Amati correlation, we can derive the jet opening 
angle distribution for the 442 GRBs with pseudo redshifts.  This is 
compared to the angle distribution of the few GRBs with measured jet 
break time and redshift (see Tab. 2 of GGL04) in Fig. 4.  
\begin{figure} 
\begin{center} 
\vspace{8cm} 
\includegraphics{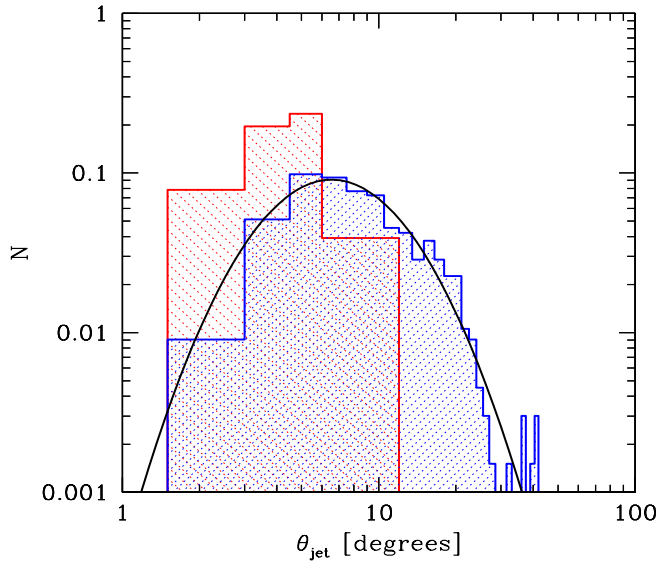} 
\caption{ Jet opening angle distribution for the sample of GRBs with
pseudo redshift (blue hatched histogram).  This has been obtained
requiring they obey the Ghirlanda correlation.  The solid black line
is the log-normal fit to this distribution, which peaks at $\sim
6.5^\circ$.  Also shown is the distribution (red hatched histogram) of
measured jet opening angles for the 17 bursts in the sample of
GGL04 of known redshifts and jet break time (Ghirlanda et
al. 2005).}
\end{center} 
\label{angoli} 
\end{figure} 
The angle distribution derived for the 442 GRBs and represented 
in Fig. 4 is described by a log-normal function:
\begin{equation} 
N(\theta_{\rm jet})\, =\, 
{1 \over \sqrt{2\pi}\,\, \theta_{\rm jet} a_{1} } 
\,\, \exp{\left[-{ (\ln\theta_{\rm jet}-a_2)^2 \over 2a_1^2} \right]} 
\end{equation}
The best fit parameters are $(a_{1}, a_{2})=(2.2, 0.57)$ which lead to
a peak value of $\sim$6.5$^\circ$.  Note that the derived distribution
is the sum of the distribution of angles for bursts with {\it any
value} of $E_{\rm peak}$, and does not pretend to be representative of
the distribution at {\it each} $E_{\rm peak}$.  Indeed, if the jet
opening angle distribution remains the same for all values of $E_{\rm
peak}$, then we should expect the same slope for the Ghirlanda and
Amati correlation.  Since this is not the case, we can conclude that
the current data indicate a trend: for smaller $E_{\rm peak}$ and then
$E_\gamma$, the average jet opening angle is larger than for large
values of $E_{\rm peak}$ and $E_\gamma$.  It is premature to conclude
whether this trend is due to selection effects or if it has a physical
origin (i.e.  smaller jet opening angles for intrinsically more
powerful bursts), and we plan to investigate in more detail this
problem in a forthcoming paper.

\section{Comparison with previous works} 
 
In the previous section we have shown that if the pseudo--redshifts 
obtained through the lag--luminosity relation are indicative of real 
redshifts, then the $E_{\rm peak}$--$E_{\rm iso}$ correlation exists 
even considering a large sample of GRBs.  We now compare this 
conclusion with the 
conclusions of NP04 and BN05.

The method  adopted by NP04  and BP05 starts  from a given  rest frame 
correlation  $E_{\rm   peak}=K  E_{\rm  iso}^{\eta}$   where  the  two 
quantities  are   related  to  the   observables   
$E_{\rm peak,obs}=E_{\rm peak}/(1+z)$   and the fluence  
$F=E_{\rm   iso}(1+z)/4\pi d_L^2(z)$.  We can form the ratio,
\begin{equation} 
{E_{\rm peak,obs}^{1/\eta}\over F} \, =\,  
K^{1/\eta}\,10^{\pm \sigma/\eta}\, {4\pi d_{L}^{2}(z)\over (1+z)^{(1+\eta)/\eta}} 
\end{equation} 
where $\sigma$ corresponds  to the scatter of data  points around the 
correlation that is being tested.   
The  RHS of the  above relation is a function of $z$ and $\eta$ only,  
and is monotonically increasing with redshift if $\eta \ge 0.75$. 
For lower values of $\eta$ it has a maximum. 

As in the case of NP04 and BP05, we take into account now the upper
limit of the ratio $E_{\rm peak,obs}^{1/\eta}/F$. This upper limit
establishes an allowance region boundary on the corresponding plane.
Furthermore, we include the scatter around the assumed correlation
increasing the previous upper limit by the factor $10^{\sigma/\eta}$
(see Eq. 9).

NP04 and BP05 are testing the $E_{\rm peak}-E_{\rm iso}$ correlation
as originally found with 9 bursts by Amati et al.  2001 , i.e.  with
slope $\eta=0.5$.  They allow for the possible scatter around this
correlation with a factor 2 in the normalization of Eq.  9.  This
corresponds, for their assumptions, to consider $\sigma \sim 0.15$.
As shown in this work (and see also GGL04) the 27 GRBs show a
dispersion with $\sigma=0.22$ in the $E_{\rm peak}-E_{\rm iso}$ plane
around a correlation with slope $\eta=0.56$ (see Eq.1). Moreover, we
consider the 3$\sigma$ consistency of our sample of bursts with this
correlation.

\begin{figure} 
\begin{center} 
\vspace{7.0cm} 
\includegraphics{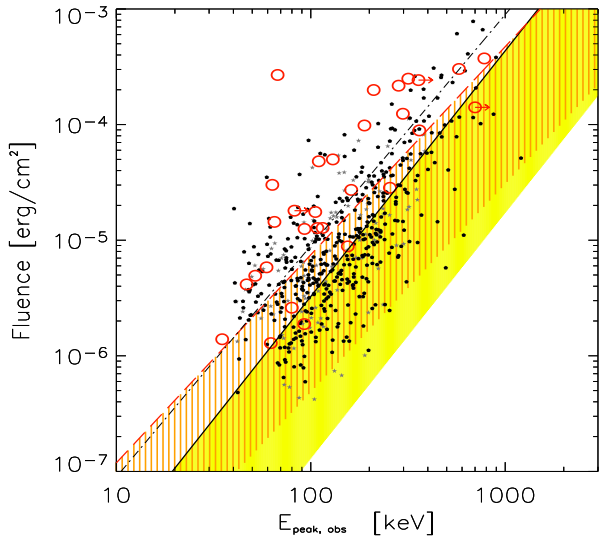} 
\caption{ Observed plane. Fluence versus peak energy of the average
spectrum for the sample of 546 GRBs selected in Sec. 2. Black dots are
the 442 GRBs which has been used in Sec. 3.  Grey stars are 104 GRBs
excluded in the analysis of Sec.  3 either for their large redshifts
or uncertain lags (see Sect.2).  Also shown are the 30 GRBs
(open red circles) with known redshift (from GGL04,
including/upper/lower limits on $E_{\rm peak}$). The dashed red line
(black solid line) represents the boundary defined by the correlation
of Eq. 1 (Eq. 6). The hatched red region (yellow filled region)
represents the $3\sigma$ shift of the corresponding boundary due to
the scatter.  Points above these boundaries are fully consistent with
the corresponding correlations.  Upper/lower limits for data points
are indicated by arrows.  The dot--dashed black line represents the
boundary of the Amati correlation adopted in BP05 and NP04.}
\end{center} 
\label{test} 
\end{figure} 

For the  above reasons we  applied the same  test adopted by  NP04 and
BP05, but using  the correlation found here, with  its proper scatter,
i.e $\sigma=0.22$.  {\it  Note also that we use the  peak energy of the
time integrated spectrum $E_{\rm peak}$}.

In Fig. 5 we report the 27 GRBs with spectroscopic redshifts
(open red circles, see. Sec. 3, plus 3 upper limits), as well as the
442 GRBs (black dots) of the sample described in Sec.  2. The dashed
red line (black solid line) represents the boundary defined by Eq. 1
(Eq. 6), while the red (yellow) hatched area shows the 3$\sigma$ shift
of this boundary related to the scatter. For a gaussian distribution
the area below the 3$\sigma$ limit is highly improbable for the
elements of the corresponding sample.  The point here is that 1.4\% of
the black dot distribution extends below the 3$\sigma$ limit of Eq. 1
correlation (see Tab. 2 for the 1$\sigma$ and 2$\sigma$ fractions).
This result makes Eq.  1 incompatible with the 442 BATSE GRB sample,
in agreement with Nakar \& Piran (2004) and Band \& Preece
(2005). This conclusion is strengthened by the fact that our allowance
region is obtained using the upper limit of
$E_{peak,obs}^{1/\eta}/F$. 

We conclude that Eq. 1 has to be considered
as a temporary estimate of the $E_{\rm peak}$-$E_{\rm iso}$ (Amati)
correlation with an associated uncertainty likely due to the still
small number of GRBs with known $z$. However, our main point is that
the BATSE GRB sample confirms the existence of an $E_{\rm
peak}$-$E_{\rm iso}$ correlation and the scatter of this correlation
($\sigma$=0.22) agrees with the scatter of the sample of the 27 GRBs
with spectroscopic redshifts (Eq. 1).

\begin{table} 
\centering 
\begin{tabular}{@{}lllll@{}} 
\hline   
& no scatter &  $1\sigma$  & $2\sigma$  & $3\sigma$ \\ 
\hline  
Eq. 1 (red dashed)  &  32\%  &  68\%  & 91.8\%  & 98.6\% \\  
Eq. 6 (black solid)       &  59\%  &  91.1\%  & 98.7\%  & 100\% \\  
\hline 
\end{tabular} 
\caption{ Consistency of the sample of 546 GRBs described in Sec. 2
with the Amati correlations as given by Eq.1 and Eq.6.  The
percentage represents the fraction of points above the lines in Fig.
5 (no scatter) or above the lower boundary of the corresponding
$\sigma$ scatter.}
\end{table}

\section{Conclusions} 

Through a large sample of 442 BATSE GRBs, with known spectral peak
energy (Y04) and pseudo redshift (BNB04), we have computed the rest
frame peak energy $E_{\rm peak}$ and the k-corrected equivalent
isotropic energy $E_{\rm iso}$. We have found that these two
quantities are correlated: the power-law fit to this correlation gives
a slope slightly flatter and an $E_{\rm iso}$ a little smaller than
the slope and the $E_{\rm iso}$ found with the 27 GRBs with
spectroscopic redshifts, respectively. The scatter of the 442 data
points around the fitting line is equal to the scatter of the 27 GRBs
with known redshift, the existence of outliers being marginal.

We have further interpreted the scatter of these data points as due to
the distribution of the jet opening angle of GRBs. By requiring that
they obey the Ghirlanda correlation, we could estimate this
distribution, which is a log-normal with an average angle of about
6.5$^{\circ}$.

Our main conclusion is that the existence of an $E_{\rm peak}$-$E_{\rm
iso}$ (Amati) correlation is confirmed, its scatter is probably rather
well known, but its definitive estimate needs a wider sample of
GRBs with spectroscopic measured redshifts.

\vspace{-0.45cm}
\section*{Acknowledgments} 
We thank Z.  Bosnjak, A. Celotti, G.  Barbiellini \& F.  Longo for
sending us their manuscript before submission. The referee J. Bonnell
is thanked for his useful comments.  We thank the italian MIUR for
funding through Cofin grant 2003020775\_002.

\end{document}